\documentclass[aps,prl,reprint,longbibliography]{revtex4-1}
\usepackage{appendix}
\usepackage{bm}
\usepackage{enumerate}
\usepackage{graphicx}% Include figure files
\usepackage{dcolumn}% Align table columns on decimal point
\usepackage{bm}% bold math
\usepackage{amssymb}
\usepackage{amsmath}
\usepackage{booktabs}
\usepackage{braket}
\usepackage{bbm}
\usepackage[caption = false]{subfig}

\usepackage{tikz}
\usetikzlibrary{patterns}
\tikzset{
    cross/.pic = {
    \draw[rotate = 45] (-#1,0) -- (#1,0);
    \draw[rotate = 45] (0,-#1) -- (0, #1);
    }
}
\usetikzlibrary{arrows.meta}
\usetikzlibrary {patterns.meta}
\pgfdeclarepattern{
      name=hatch,
      parameters={\hatchsize,\hatchangle,\hatchlinewidth},
      bottom left={\pgfpoint{-.1pt}{-.1pt}},
      top right={\pgfpoint{\hatchsize+.1pt}{\hatchsize+.1pt}},
      tile size={\pgfpoint{\hatchsize}{\hatchsize}},
      tile transformation={\pgftransformrotate{\hatchangle}},
      code={
        \pgfsetlinewidth{\hatchlinewidth}
        \pgfpathmoveto{\pgfpoint{-.1pt}{-.1pt}}
        \pgfpathlineto{\pgfpoint{\hatchsize+.1pt}{\hatchsize+.1pt}}
        \pgfpathmoveto{\pgfpoint{-.1pt}{\hatchsize+.1pt}}
        \pgfpathlineto{\pgfpoint{\hatchsize+.1pt}{-.1pt}}
        \pgfusepath{stroke}
      }
    }

\tikzset{
      hatch size/.store in=\hatchsize,
      hatch angle/.store in=\hatchangle,
      hatch line width/.store in=\hatchlinewidth,
      hatch size=3pt,
      hatch angle=0pt,
      hatch line width=.5pt,
    }

\newcommand{\ee}{{\rm e}}
\newcommand{\ii}{{\rm i}}
\newcommand{\rr}{\bm r}

\newcommand{\adag}{a^{\dagger}}
\newcommand{\R}{\mathbb R}

\newcommand{\ipcurved}[2]{\left\langle #1,#2\right\rangle_g}

\tikzset{every picture/.style={line width=0.75pt}} %set default line width to 0.75pt        

\begin{document}

\preprint{APS/123-QED}

\title{Designing Hamiltonians with Exactly Flat Higher-Vortexable Bands}% Force line breaks with \\

\author{Daniel Spasic-Mlacak  and Nigel R. Cooper}

\affiliation{
T.C.M Group, Cavendish Laboratory, University of Cambridge, J.J. Thomson Avenue, Cambridge CB3 0US, United Kingdom \looseness=-1}
\date{\today}

\begin{abstract}
The zero modes of the Aharonov--Casher (AC) Hamiltonian form an exactly flat, vortexable Chern band in any nonuniform magnetic field, analogous to the lowest Landau level. We ask which local Hamiltonians host exactly flat \emph{higher}-vortexable bands. For a scalar Hamiltonian, an exactly flat first-vortexable band requires a mass proportional to the local field; the Dirac analogue has the same structure. For periodic Bernal bilayer graphene with nonzero mean flux, the exact zero-mode pair is genuinely first-vortexable if and only if the interlayer hopping has nonzero mean; zero-mean hopping instead gives a decomposable pair. Multiorbital models yield exact higher-vortexable chains to arbitrary order, while remaining first order in momentum.
\end{abstract}

\maketitle

Landau levels of a charged particle in a uniform magnetic field are exactly flat, topological bands with a rigid analytic structure that underlies much of quantum Hall physics. Aharonov and Casher showed that part of this structure survives in a nonuniform field: the zero modes of the two-dimensional Pauli Hamiltonian are spin polarized and retain a holomorphic structure~\cite{aharonov1979ground}. The resulting Aharonov--Casher (AC) band is thus an analogue of the lowest Landau level (LLL), with an envelope defined by the local field $B(\mathbf r)$.

In the search for fractional Chern insulators, i.e. lattice generalizations of the fractional quantum Hall effect ~\cite{regnault2011fractional,parameswaran2013fractional,cai2023signatures,park2023observation,lu2024fractional}, \emph{vortexability} has been identified as the property that makes a band LLL-like: closure under multiplication by the complex coordinate \cite{ledwith2023vortexability}. \emph{Higher vortexability} is the corresponding property for higher Landau levels defined on an $n$-band subspace: it requires multiplicative closure of the full $n$-band complex, together with indecomposability into smaller sub-complexes that are themselves vortexable~\cite{fujimoto2025higher, liu2025theory}. It was shown that the first-vortexable complex of two bands (analogous to the lowest and first excited Landau levels) can be approximately realized in models for strained Bernal graphene with zero-mean periodic pseudomagnetic field. Here we address the design problem: \textit{which local Hamiltonians have exactly flat, higher-vortexable bands? Furthermore, can higher vortexable bands survive a nonuniform magnetic field analogous to the lowest AC band?}

We address these questions for four classes of Hamiltonians. First, we extend the AC Hamiltonian to have, besides the AC band, an exactly flat first-vortexable band at higher energy. Requiring this fixes the kinetic coefficient to be inversely proportional to the local field, and for local second-order scalar Hamiltonians it is the only choice. Second, we give the relativistic version, a two-component Dirac Hamiltonian with the same structure. Third, we show that the zero modes of chiral Bernal-stacked bilayer graphene remain exactly degenerate for any $B(\rr)$ and $\gamma(\rr)$, generalizing results for the quadratic-in-momentum effective model~\cite{katsnelson2008zero,kailasvuori2009pedestrian}, and form zeroth- and first-vortexable bands if and only if $\braket{\gamma}\neq0$; at zero mean flux the periodic Bernal AC complex is obstructed from being genuinely first-vortexable. Fourth, we generalize to first-order chiral models with $n+1$ orbitals per chirality, whose exact zero modes are vortexable up to any order $n$, with Bernal bilayer graphene as the $n=1$ case and chiral rhombohedral multilayers as a special case. Higher vortexability arises from the multiorbital structure, with the Hamiltonian remaining first order in momentum.

\begin{figure}[t]
\centering
\includegraphics[width=3.42in]{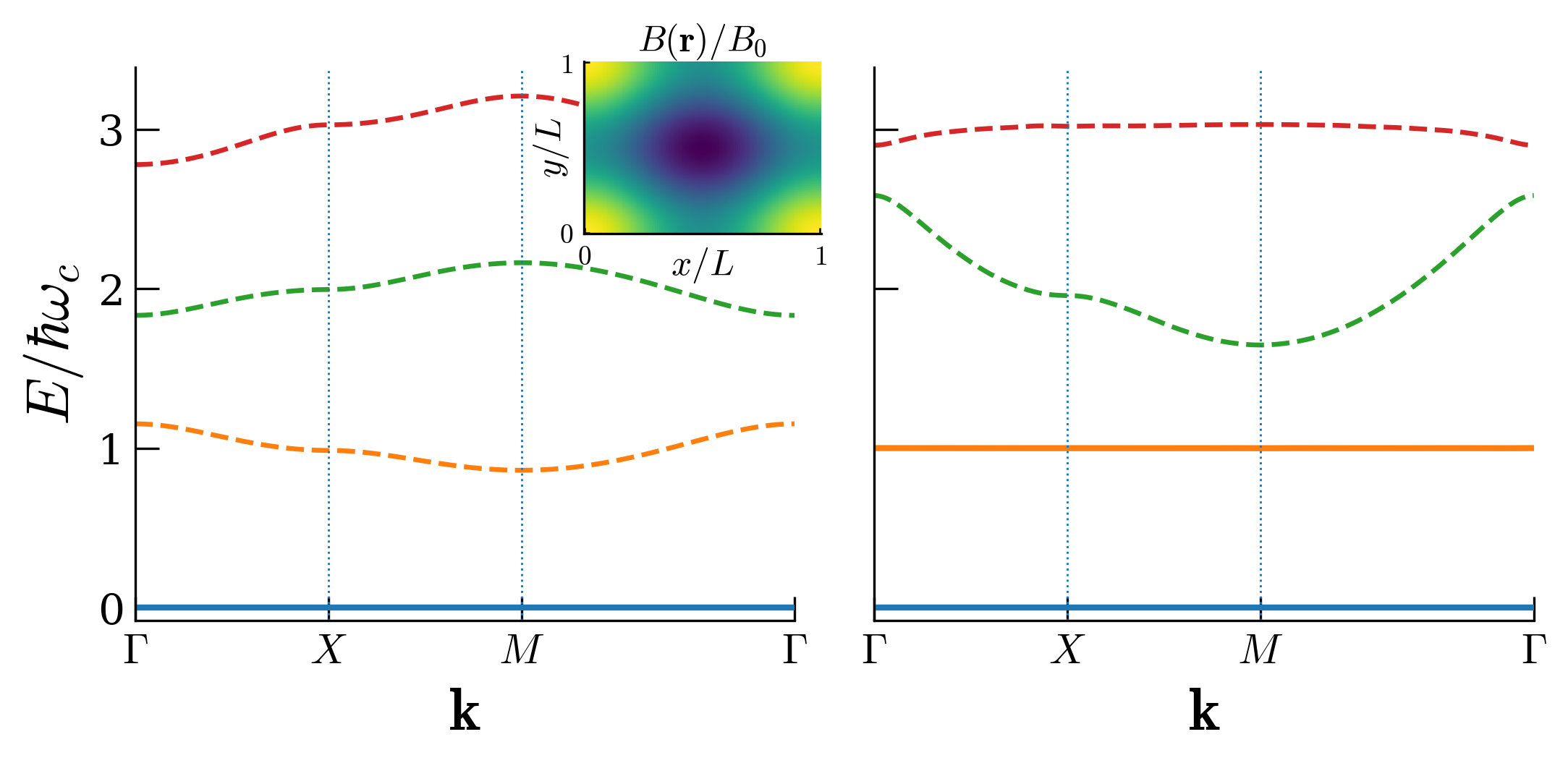}\
\caption{Band structures for the magnetic field $B(\mathbf r)=B_{0}[1+\frac{\epsilon}{2}(\cos(Gx)+\cos(Gy))]$ with $\epsilon=0.7$ and one flux quantum per magnetic unit cell ($G=2\pi/L$, $L^{2}=2\pi \ell_{B_{0}}^{2}$); the field profile is shown inset. (left) For the conventional Aharonov--Casher Hamiltonian $H_{\rm AC}=\hbar \omega_{c} a^\dagger a$ with $\omega_{c} = B_{0}/M$, only the lowest band is exactly flat, while the excited bands are generically dispersive. (right) For the Hamiltonian $H_{\mathrm{AC}}^{\star}=\hbar \omega_{c}a^\dagger \frac{B_{0}}{B(\rr)}a$, the first excited band is also exactly flat, producing a two-band first-vortexable complex. Solid lines denote the protected flat bands, and dashed curves the remaining dispersive bands.  }
\label{Figure1_highervortex}
\end{figure}

To formulate the continuum problem, consider a charged particle with mass $M$ in two dimensions, subject to an external nonuniform magnetic field (and associated vector potential ${\bm A}$) $B(\rr) = \partial_xA_y-\partial_yA_x$. The same formalism also applies to effective or synthetic magnetic fields, such as those arising in moiré systems, including twisted bilayer graphene~\cite{ledwith2020fractional,wang2021exact,shi2024adiabatic}, or cold atomic gases~\cite{dalibard2011colloquium,cooper2019topological}; we discuss this connection further in the Supplementary Material.
We assume $B>0$ in the AC and Dirac examples, where $B^{-1}$ is used, but require no such assumption for the (multiorbital) Bernal cases. We choose a reference field $B_{0}$ [e.g. the average $\braket{B(\textbf{r})}$] and magnetic length $\ell_{B_0} = \sqrt{1/B_0}$ and work with units $\hbar = q = \ell_{B_{0}} = 1$, where $q$ is the charge of the particle. 
The AC Pauli Hamiltonian for the spin sector aligned with the magnetic field is then given by\cite{aharonov1979ground}
\begin{equation}
H_{\mathrm{AC}} = \frac{[-\ii \nabla_{\rr} - \bm A (\rr)]^{2}}{2M} - \frac{B(\rr)}{2M}. 
\label{eq:AChamiltonian}
\end{equation}
Using the canonical momentum operators $\Pi_{i} = -\ii \partial_{i} - A_{i} (\rr)$, we define the analogue of the Landau-level annihilation operator in a nonuniform magnetic field as
\begin{equation}
a=\frac{1}{\sqrt{2}}(\Pi_x + \ii \Pi_y) = \frac{1}{\sqrt 2} ( - 2\ii \partial_{\bar{z}}  - A),
\end{equation}
where $z = x+\ii y$, $\partial_{\bar{z}} = \frac{1}{2}(\partial_{x}+\ii \partial_{y})$, and $A=A_{x}+\ii A_y$. 
For a nonuniform field, these operators factorise the Pauli Hamiltonian into $H_{\mathrm{AC}} = \frac{1}{M} \adag a $, but do not form a canonical oscillator algebra. What prevents a ladder of excited states is the spatial dependence of the commutator
\begin{equation}[a,a^\dagger]=B(\rr).
\label{eq:comm}
\end{equation}
The zero modes of \eqref{eq:AChamiltonian} obey $a\phi_0=0$, and their solutions are given by 
\begin{equation}
\phi_{0}(\rr) = f(z) \ee^{-K(\rr)},
\label{eq:ACzeromodes}
\end{equation}
where $B(\rr) =  4 \partial_{z}\partial_{\bar{z}}K(\rr)$, underlying the Aharonov--Casher zero-mode theorem \cite{aharonov1979ground, erdHos2002pauli}.

\textit{Generalizing Aharonov--Casher.--}
The AC band is the lowest vortexable band. For a Chern-number $C=1$ first-vortexable band, Ref.~\cite{fujimoto2025higher} identifies the canonical single-particle form
\begin{equation}
\bm\phi_1(\rr)
=
Q_0\left[
\left(
\bar z f(z)-2f'(z)
\right)\bm{\mathcal N}_0(\rr)
+
f(z)\bm{\mathcal N}_1(\rr)
\right],
\label{eq:firstvortexstate}
\end{equation}
where $Q_0=1-P_0$ projects orthogonally to the lowest-vortexable band. We restrict to this canonical form; Ref.~\cite{fujimoto2025higher} left open whether it exhausts all first-vortexable states. For the scalar AC problem it may be written locally as
\begin{equation}
\tilde{\phi}_{1}
=
Q_0\left[a^\dagger+h(\rr)\right]\phi_0 ,
\label{eq:generalstate}
\end{equation}
where $h(\rr)$ is a state-independent multiplication operator. Any local scalar Hamiltonian of momentum order at most two that annihilates every AC zero mode, $H\phi_0=0$, takes the form
\begin{equation}
H_{w}=a^\dagger w(\rr)a,
\label{eq:H_w}
\end{equation}
with $w(\rr)$ real (Supplementary Material). Every such Hamiltonian has the AC band at zero energy, with the conventional AC Hamiltonian~\eqref{eq:AChamiltonian} recovered for constant $w$. 

To make contact with the canonical first-vortexable form \eqref{eq:firstvortexstate}, we separate the uniform and nonuniform parts of the AC envelope, $K(\rr)=\frac{|z|^2}{4}+\delta K(\rr)$,
where $\delta K(\rr)$ describes the zero-average modulation of the magnetic field (correspondingly,
$B(\rr)
=1+4\partial_z\partial_{\bar z}\delta K(\rr)$ in our choice of units). The state $a^{\dagger}\phi_{0}$ then has the form of Eq.~\eqref{eq:firstvortexstate}, up to normalization, with $\mathcal{N}_{0}(\rr)=\ee^{-K(\rr)}$ and $\mathcal{N}_{1}(\rr)=4\partial_{z}(\delta K(\rr))\ee^{-K(\rr)}$, and $Q_{0}$ acts trivially on it. Using Eq.~\eqref{eq:comm},
\begin{equation}
H_w\,a^\dagger\phi_0 = a^\dagger\, w(\rr)B(\rr)\,\phi_0.
\label{eq:a_dagger_argument}
\end{equation}
This is proportional to $a^\dagger\phi_0$ for all $f$ only if $wB=E_1$ is constant. For $w$ constant this fails when $B$ is nonuniform: the standard AC Hamiltonian protects the lowest band but not its first-Landau-level partner. Taking $w=E_1B^{-1}$ instead gives
\begin{equation}
H^{\star}_{\mathrm{AC}} = E_{1} \adag B(\rr)^{-1} a ,
\label{eq:firstvortexableAC}
\end{equation}
which has the AC band at $E=0$ and an exactly flat first-vortexable band $a^\dagger\phi_0$ at $E=E_1$, for any smooth $B(\rr)$ that is nowhere zero (Fig.~\ref{Figure1_highervortex}).

We find that this choice is unique: requiring $H_w\tilde\phi_1=E_1\tilde\phi_1$ for every $\tilde{\phi}_1$ of the form Eq.~\eqref{eq:generalstate} forces $w(\rr)B(\rr)=E_1$ (Supplementary Material). Since the argument uses only the local form of the AC zero modes~\eqref{eq:ACzeromodes}, it holds independently of boundary conditions.

Physically, first vortexability requires the local effective mass to follow the field, $M_{\rm eff}(\rr)\propto B(\rr)$. It also fixes the operator ordering and hence the mass-gradient coupling. This ordering differs from the one obtained by putting the spatial dependence into a curved-space metric\cite{iengo1994quantum,estienne2023ideal}, so $H^\star_{\rm AC}$ is not the ordinary Pauli Hamiltonian in a uniform field on a curved surface. The comparison with position-dependent-mass and curved-space Hamiltonians is in the Supplementary Material.

\textit{Dirac Hamiltonian.--}
The same question can be asked for first-order dynamics. For a local two-component chiral Hamiltonian, the analogue of $H^\star_{\rm AC}$ is
\begin{equation}
 H_{\mathrm D}^{\star} = \Delta
 \begin{pmatrix}
 0&a^\dagger B^{-1/2}\\
 B^{-1/2}a&0
 \end{pmatrix}.
 \label{eq:2x2Dirac}
\end{equation}
The spectrum of $H_{\mathrm D}^{\star}$ follows directly from the AC zero modes. For every $\phi_0\in\ker a$, the spinor $\bm\phi_0=(\phi_0,0)^{T}$ is a zero mode. The spinors $\bm\phi_1=(a^\dagger\phi_0,\pm B^{1/2}\phi_0)^{T}$ are eigenstates at $E=\pm\Delta$, because $(B^{-1/2}a)a^\dagger\phi_0=B^{1/2}\phi_0$ and $(a^\dagger B^{-1/2})B^{1/2}\phi_0=a^\dagger\phi_0$. As $\phi_0$ ranges over $\ker a$, these give three exactly flat bands, at $E=0$ and $E=\pm\Delta$.

The bands at $\pm\Delta$ are first-vortexable partners of the zero-energy band. Since $[z,a^\dagger]=\lambda$ and $z$ commutes with $B^{1/2}$, multiplication by $z$ gives $z\bm\phi_1[\phi_0]=\bm\phi_1[z\phi_0]+\lambda\bm\phi_0[\phi_0]$. Vortex attachment thus keeps a state within its own band, up to a component in the zero-energy band. The finite-energy spinors therefore have the multicomponent first-vortexable form of Eq.~\eqref{eq:firstvortexstate}, with $\bm{\mathcal N}_0=(\ee^{-K},0)^{T}$ and $\bm{\mathcal N}_1=\ee^{-K}\bigl(4\partial_z\delta K,\mp\ii\sqrt2B^{1/2}\bigr)^{T}$, up to normalization. Each pair of bands, at $E=0$ and $E=+\Delta$ or at $E=0$ and $E=-\Delta$, is therefore a zeroth- and first-vortexable complex.

The square of $H_{\mathrm D}^{\star}$ is block diagonal. Its upper block, $\Delta^2a^\dagger B^{-1}a=(\Delta^2/E_1)H^{\star}_{\rm AC}$, is the scalar Hamiltonian found above. In $(H_{\mathrm D}^{\star})^2$ the two bands at $\pm\Delta$ become degenerate at $\Delta^2$, and their common upper component $a^\dagger\phi_0$ is the first-vortexable band of $H^{\star}_{\rm AC}$. Although $H^\star_{\rm AC}$ is not itself the Pauli Hamiltonian on a curved surface, the other block of $(H_{\mathrm D}^{\star})^2$ is $\Delta^2B^{-1/2}aa^\dagger B^{-1/2}=\Delta^2\bigl(1+B^{-1/2}a^\dagger aB^{-1/2}\bigr)$, using Eq.~\eqref{eq:comm}. Up to this constant shift, $B^{-1/2}a^\dagger aB^{-1/2}$ is the Hamiltonian of a charged particle in a uniform field on a curved surface with conformal metric $\sqrt g=B(\rr)$, written in the flat measure (Supplementary Material). Its lowest Landau level, $B^{1/2}\ker a$, is flat and vortexable~\cite{estienne2023ideal}, and it supplies the lower components of $\bm\phi_1$. Higher Landau levels are degenerate on surfaces of constant curvature~\cite{iengo1994quantum}, but for this Hamiltonian the analysis of symmetric scalar dressings in the Supplementary Material shows that an exactly flat first-vortexable band requires $B$ to be uniform. Since the two blocks share their nonzero spectrum, the same obstruction prevents $H^\star_{\rm AC}$ from having flat bands beyond the first-vortexable one.

As in the scalar case, $H_{\mathrm D}^{\star}$ is essentially the unique Hamiltonian of this type. Consider a general local chiral Hamiltonian $\Delta\bigl(\begin{smallmatrix}0&D_0^\dagger\\ D_0&0\end{smallmatrix}\bigr)$, with $D_0$ a scalar first-order operator. For $(\phi_0,0)^T$ to be a zero mode for every $\phi_0\in\ker a$, $D_0$ must annihilate every AC zero mode, which forces $D_0=\mu(\rr)a$ (Supplementary Material). This gives the family
\begin{equation}
H_{\mu}=\Delta\begin{pmatrix}0&\adag \mu^{*}(\rr)\\ \mu(\rr) a&0\end{pmatrix}.
\label{eq:mudiracfamily}
\end{equation}
Further requiring an exactly flat finite-energy first-vortexable partner of the form $(a^\dagger\phi_0,c(\rr)\phi_0)^T$ yields
$|\mu(\rr)|\propto B^{-1/2}(\rr)$. The family of Hamiltonians in Eq.~\eqref{eq:mudiracfamily} corresponding to first-vortexable states are then exactly those of Hamiltonian ~\eqref{eq:2x2Dirac}, with an additional position dependent phase $\ee^{\ii\theta(\rr)}$, where $\mu(\rr)\propto \ee^{\ii\theta(\rr)} B^{-1/2}(\rr)$ and $c(\rr) = \pm\ee^{\ii\theta(\rr)} B^{1/2}(\rr)$. This phase can be removed locally by the transformation $U=\mathrm{diag}(1,\ee^{\ii\theta(\rr)})$, provided it has no topologically non-trivial global phase winding, so Eq.~\eqref{eq:2x2Dirac} is unique up to this phase. 
In any case, the phase drops out of the square of $H^{\star}_{\mathrm{D}}$, whose upper block is always proportional to $a^\dagger B^{-1}a$.
A constant Dirac mass term $m_{\mathrm{D}}\sigma_{z}$ preserves the construction, moving the zero-mode band to $E=m_{\mathrm{D}}$ and shifting the finite-energy pair to $E_\pm=\pm\sqrt{m_{\mathrm{D}}^2+\Delta^2}$, while leaving the first-vortexable structure unchanged up to a fixed spinor mixing angle.

Neither the scalar second-order class nor the Dirac class can give higher vortexability beyond first order. For example, for the state $\phi_2=(\adag)^{2}\phi_0$, we have
$a(\adag)^2\phi_0=2B\adag\phi_0+[\adag,B]\phi_0$.
More generally,
\begin{equation}
[a,(\adag)^n]=\sum_{\ell=0}^{n-1}(\adag)^\ell B (\adag)^{n-1-\ell}.
 \label{eq:highervortexobstruction}
\end{equation}
For nonconstant $B$, normal ordering generates $[\adag,B]$, $[\adag,[\adag,B]],\ldots$ multiplying independent holomorphic derivatives of $f$. A single function $w(\rr)$ in $H_w$ cannot in general cancel all of these terms for $n\ge2$. In a uniform field the nested commutators vanish and the usual Landau-level ladder is recovered. 
As we show below, exact higher vortexability can be achieved in models that go beyond the classes considered so far.

\textit{Bernal-stacked bilayer graphene.--}
The low-energy two-band description of Bernal-stacked bilayer graphene is quadratic in momentum, but the underlying chiral Hamiltonian is first-order, with the extra structure in the orbital degrees of freedom:
\begin{equation}
H_{\rm B}\sim
\begin{pmatrix}0&D_{1,\gamma}^\dagger\\D_{1,\gamma}&0\end{pmatrix},
 \quad
 D_{1,\gamma} =  \begin{pmatrix}a&\gamma\\0&a\end{pmatrix}.
 \label{eq:Bernal}
\end{equation}
Here $\gamma$ is the dimensionless dimer hopping, which for now we consider to be spatially uniform.
In a uniform field $B(\rr)=B_{0}$, solving $D_{1,\gamma}\Psi=0$ gives one family $(\phi_0,0)^T$ ($a\phi_0=0$), and a second family whose upper component contains $a^\dagger\phi_0$. Eliminating the auxiliary component gives $a^2\phi_1=0$, the familiar $n=0,1$ orbital doublet of ideal Bernal bilayer graphene.
Thus, in a uniform magnetic field, this model contains degenerate $E=0$ bands which are zeroth- and first-vortexable~\cite{fujimoto2025higher}.

For spatially varying $B(\rr)$ and $\gamma(\rr)$, the zero modes remain exactly degenerate. Since $\ker a^\dagger=\{0\}$ for positive flux and $D_{1,\gamma}$ is upper triangular with $a$ on the diagonal, solving $D_{1,\gamma}\Psi=0$ row by row gives $\dim\ker D_{1,\gamma}=2N_\phi$, generalizing Refs.~\cite{katsnelson2008zero,kailasvuori2009pedestrian} from the quadratic-in-momentum effective model to the linear-in-momentum model. Since $[D_{1,\gamma},z]=0$, $\ker D_{1,\gamma}$ is closed under $z$~\cite{ledwith2023vortexability}. It remains to determine whether the two bands are indecomposable~\cite{fujimoto2025higher}. 

Writing the second family in the local form $\bigl((a^\dagger+h)\phi, c\phi\bigr)^T$ requires
\begin{equation}
\gamma=-B/c-[a,h].
\label{eq:Bernalcond}
\end{equation}
Multiplication by $z$ maps this second family, normalized as $\bigl(c^{-1}(a^\dagger+h)\phi,\phi\bigr)^T$, into itself plus $\lambda/c$ times the lowest-vortexable family $(\phi_0,0)^{T}$, where $\lambda\equiv[z,a^\dagger]$. For periodic $h$ and $\braket{B}\neq0$, averaging Eq.~\eqref{eq:Bernalcond} gives $\lambda/c=-\lambda\braket{\gamma}/\braket{B}$. The two zero-mode bands are therefore lowest- and first-vortexable if and only if $\braket{\gamma}\neq0$. If $\braket{\gamma}=0$, then $\gamma=-[a,h]$, the family $(h\phi_0,\phi_0)^T$ is itself closed under $z$, and the complex decomposes.

Thus, for nonuniform $B(\rr)$ with $\braket{B}\neq0$, the model~\eqref{eq:Bernal} hosts both zeroth- and first-vortexable bands for any $\gamma(\rr)$ with $\braket{\gamma}\neq0$. In the scalar and Dirac models, first vortexability fixed $w(\rr)$ and $\mu(\rr)$ uniquely. Here the extra orbital removes that constraint, because the state-independent function $h$ takes up the inhomogeneity. Taking $\gamma=\gamma_0$ constant, Eq.~\eqref{eq:Bernalcond} becomes $[a,h]=-B/c-\gamma_0$, which can be solved locally for $h$ for any smooth $B(\rr)$. For nonzero mean field, the periodic solvability condition fixes only the average coupling, $\gamma_0=-\braket{B}/c$, so Bernal-stacked bilayer graphene with uniform interlayer hopping can retain exact lowest- and first-vortexable zero-mode bands even in a nonuniform magnetic field. 
Setting $h=0$ gives $\gamma\propto B$, but nothing requires this: for any $\gamma(\rr)$ (with $\braket{\gamma}\neq 0$) and any $B(\rr)$ we can choose $h$ to get a solution. 

At zero mean flux, there is no genuine periodic Bernal AC complex. If $\braket{B}=0$ but $\braket{\gamma}\neq0$, the spatial average of Eq.~\eqref{eq:Bernalcond} cannot hold. If $\braket{B}=\braket{\gamma}=0$, a periodic $h_0$ satisfying $\gamma = -[a,h_{0}]$ exists and explicitly decomposes the zero-mode space into two separately vortexable families (see Supplementary Material). This is consistent with the periodically strained Bernal model of Ref.~\cite{fujimoto2025higher}: there the pseudomagnetic field has zero spatial average and the Bernal coupling is constant. The resulting lowest- and first-vortexable states are hence only approximate. Our condition identifies the obstruction to promoting such a constant-coupling zero-mean-field construction to an exact AC realization.

\textit{Multiorbital generalization.--}
The Bernal construction extends to arbitrary vortexability by including additional orbitals. For an \((n+1)\)-orbital system, consider an upper-triangular operator with \(a\) on the diagonal and local multiplication operators above it. Such operators retain \((n+1)N_\phi\) zero modes, while the interorbital couplings determine how these modes are connected under vortex multiplication.

To construct a single chain, consider the family
$\bm{\Phi}_n=
\bigl((a^\dagger)^n\phi_0,(a^\dagger)^{n-1}\phi_0,\ldots,a^\dagger\phi_0,\phi_0\bigr)^T$.
There exists a local, upper-triangular operator \(D_n\) that annihilates this family, with matrix elements
\begin{equation}
(D_n)_{ij}=a\delta_{ij}
-\binom{n-i}{n-j}\mathrm{ad}_{a^\dagger}^{(j-i-1)}(B)\,\mathbbm{1}_{i<j},
\label{eq:D_n}
\end{equation}
where $i,j=0,\ldots,n$, $\mathrm{ad}_{a^\dagger}^{(m)}(B)$ denotes $m$ nested commutators with $a^\dagger$ and $\mathbbm{1}_{i<j}$ is only non-zero when $i<j$. Thus, all interorbital couplings depend only on the magnetic field and its derivatives, while the differential order remains one. For $n=1$, this reduces to the Bernal operator $D_{1,\gamma}$ in \eqref{eq:Bernal} with $\gamma = -B$ ($c=1$, $h=0$).

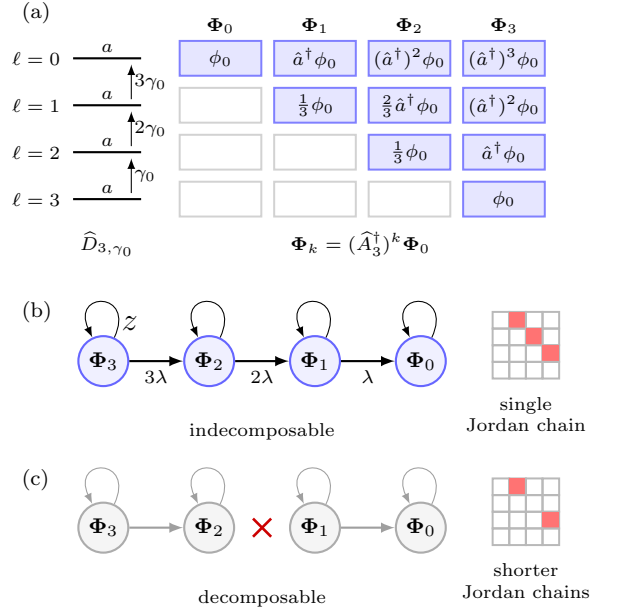
\begin{figure}[t]
\centering
\begin{tikzpicture}[
    >=latex,
    font=\footnotesize,
    fam/.style={
        circle,
        draw=blue!60,
        thick,
        minimum size=6.6mm,
        inner sep=0pt,
        fill=blue!6
    },
    badfam/.style={
        circle,
        draw=gray!65,
        thick,
        minimum size=6.6mm,
        inner sep=0pt,
        fill=gray!8
    },
    zloop/.style={
        ->,
        looseness=6,
        thin
    },
    support/.style={
        draw=blue!50,
        fill=blue!10,
        inner sep=1pt,
        font=\scriptsize,
        minimum width=1.08cm,
        minimum height=0.46cm
    },
    emptysupport/.style={
        draw=gray!35,
        inner sep=1pt,
        minimum width=1.08cm,
        minimum height=0.46cm
    }
]

% ============================================================
% (a) Orbital structure and exact Phi_k components
% ============================================================

\node[anchor=west] at (-0.55,3.60) {(a) };

% Chiral orbital stack
\foreach \l in {0,...,3}{
    \pgfmathsetmacro{\y}{3.00-0.62*\l}

    \draw[thick]
        (0.25,\y) -- (1.15,\y);

    \node[
        anchor=east,
        font=\scriptsize
    ] at (0.20,\y)
        {$\ell=\l$};

    \node[
        font=\scriptsize
    ] at (0.70,\y+0.14)
        {$a$};
}

% Inter-orbital couplings
\foreach \l in {0,1}{
    \pgfmathsetmacro{\ya}{3.00-0.62*\l-0.08}
    \pgfmathsetmacro{\yb}{3.00-0.62*(\l+1)+0.08}
    \pgfmathtruncatemacro{\hop}{3-\l}

    \draw[<-,thin]
        (1.02,\ya) -- (1.02,\yb)
        node[midway,right,font=\scriptsize,inner sep=1pt]
        {$\hop\gamma_0$};
}

\pgfmathsetmacro{\ya}{3.00-0.62*2-0.08}
\pgfmathsetmacro{\yb}{3.00-0.62*(2+1)+0.08}

\draw[<-,thin]
        (1.02,\ya) -- (1.02,\yb)
        node[midway,right,font=\scriptsize,inner sep=1pt]
        {$\gamma_0$};

\node[
    font=\scriptsize
] at (0.70,0.55)
    {$\widehat D_{3,\gamma_0}$};

% ------------------------------------------------------------
% Fixed coordinates for perfectly aligned component grid
% ------------------------------------------------------------

\def\xA{2.20}
\def\xB{3.45}
\def\xC{4.70}
\def\xD{5.95}

\def\yA{3.00}
\def\yB{2.38}
\def\yC{1.76}
\def\yD{1.14}

% Column labels
\node[font=\scriptsize] at (\xA,3.43) {$\bm\Phi_0$};
\node[font=\scriptsize] at (\xB,3.43) {$\bm\Phi_1$};
\node[font=\scriptsize] at (\xC,3.43) {$\bm\Phi_2$};
\node[font=\scriptsize] at (\xD,3.43) {$\bm\Phi_3$};

% ------------------------------------------------------------
% Phi_0
% ------------------------------------------------------------

\node[support]
    at (\xA,\yA)
    {$\phi_0$};

\node[emptysupport]
    at (\xA,\yB) {};

\node[emptysupport]
    at (\xA,\yC) {};

\node[emptysupport]
    at (\xA,\yD) {};

% ------------------------------------------------------------
% Phi_1
% ------------------------------------------------------------

\node[support]
    at (\xB,\yA)
    {$\hat a^\dagger\phi_0$};

\node[support]
    at (\xB,\yB)
    {$\frac{1}{3}\phi_0$};

\node[emptysupport]
    at (\xB,\yC) {};

\node[emptysupport]
    at (\xB,\yD) {};

% ------------------------------------------------------------
% Phi_2
% ------------------------------------------------------------

\node[support]
    at (\xC,\yA)
    {$(\hat a^\dagger)^2\phi_0$};

\node[support]
    at (\xC,\yB)
    {$\frac{2}{3}\hat a^\dagger\phi_0$};

\node[support]
    at (\xC,\yC)
    {$\frac{1}{3}\phi_0$};

\node[emptysupport]
    at (\xC,\yD) {};

% ------------------------------------------------------------
% Phi_3
% ------------------------------------------------------------

\node[support]
    at (\xD,\yA)
    {$(\hat a^\dagger)^3\phi_0$};

\node[support]
    at (\xD,\yB)
    {$(\hat a^\dagger)^2\phi_0$};

\node[support]
    at (\xD,\yC)
    {$\hat a^\dagger\phi_0$};

\node[support]
    at (\xD,\yD)
    {$\phi_0$};

\node[
    font=\scriptsize
] at (4.05,0.55)
    {$\bm\Phi_k=(\widehat A_3^\dagger)^k\bm\Phi_0$};

% ============================================================
% (b) Indecomposable vortexability chain
% ============================================================

\begin{scope}[yshift=-0.55cm]

\node[anchor=west]
    at (-0.55,0.20)
    {(b)};

% Equally spaced family nodes
\node[fam] (b3) at (0.65,-0.45) {$\bm\Phi_3$};
\node[fam] (b2) at (2.05,-0.45) {$\bm\Phi_2$};
\node[fam] (b1) at (3.45,-0.45) {$\bm\Phi_1$};
\node[fam] (b0) at (4.85,-0.45) {$\bm\Phi_0$};

% Lowering arrows
\draw[->,thick]
    (b3) -- node[
        below,
        font=\scriptsize
    ] {$3\lambda$} (b2);

\draw[->,thick]
    (b2) -- node[
        below,
        font=\scriptsize
    ] {$2\lambda$} (b1);

\draw[->,thick]
    (b1) -- node[
        below,
        font=\scriptsize
    ] {$\lambda$} (b0);

% z preserves each family
\foreach \n in {b3,b2,b1,b0}{
    \draw[zloop]
        (\n) to[out=65,in=115] (\n);
}

\node[
    font=\scriptsize
] at (1.0,0.05)
    {\large$z$};

% ------------------------------------------------------------
% Small Jordan-block icon
% ------------------------------------------------------------

\foreach \i in {0,...,3}{
    \foreach \j in {0,...,3}{

        \pgfmathsetmacro{\xx}{5.8+0.22*\j}
        \pgfmathsetmacro{\yy}{0.2-0.22*\i}
        \pgfmathtruncatemacro{\d}{\j-\i}

        \draw[gray!55]
            (\xx,\yy-0.22)
            rectangle
            (\xx+0.22,\yy);

        \ifnum\d=1
            \fill[red!55]
                (\xx,\yy-0.22)
                rectangle
                (\xx+0.22,\yy);
        \fi
    }
}

\node[
    font=\scriptsize,
    align=center
] at (6.24,-1.15)
    {single\\Jordan chain};

\node[
    font=\scriptsize
] at (2.75,-1.38)
    {indecomposable};

% ============================================================
% (c) Decomposable alternative
% ============================================================

\node[anchor=west]
    at (-0.55,-2.02)
    {(c)};

% Same overall spacing as panel (b)
\node[badfam] (c3) at (0.65,-2.65) {$\bm\Phi_3$};
\node[badfam] (c2) at (2.05,-2.65) {$\bm\Phi_2$};

\node[badfam] (c1) at (3.45,-2.65) {$\bm\Phi_1$};
\node[badfam] (c0) at (4.85,-2.65) {$\bm\Phi_0$};

% First short chain
\draw[->,thick,gray!75]
    (c3) -- (c2);

% Second short chain
\draw[->,thick,gray!75]
    (c1) -- (c0);

% z loops
\foreach \n in {c3,c2,c1,c0}{
    \draw[
        ->,
        looseness=6,
        thin,
        gray!75
    ]
    (\n) to[out=65,in=115] (\n);
}

% Visual break between chains
\draw[
    red!80!black,
    line width=1.0pt
]
    (2.63,-2.77) -- (2.87,-2.53);

\draw[
    red!80!black,
    line width=1.0pt
]
    (2.63,-2.53) -- (2.87,-2.77);

% ------------------------------------------------------------
% Block-diagonal Jordan icon
% ------------------------------------------------------------

\foreach \i in {0,...,3}{
    \foreach \j in {0,...,3}{

        \pgfmathsetmacro{\xx}{5.80+0.22*\j}
        \pgfmathsetmacro{\yy}{-2.0-0.22*\i}

        \draw[gray!55]
            (\xx,\yy-0.22)
            rectangle
            (\xx+0.22,\yy);

        % Two length-2 Jordan blocks
        \ifnum\i=0
            \ifnum\j=1
                \fill[red!55]
                    (\xx,\yy-0.22)
                    rectangle
                    (\xx+0.22,\yy);
            \fi
        \fi

        \ifnum\i=2
            \ifnum\j=3
                \fill[red!55]
                    (\xx,\yy-0.22)
                    rectangle
                    (\xx+0.22,\yy);
            \fi
        \fi
    }
}

\node[
    font=\scriptsize,
    align=center
] at (6.24,-3.35)
    {shorter\\Jordan chains};

\node[
    font=\scriptsize
] at (2.75,-3.58)
    {decomposable};

\end{scope}

\end{tikzpicture}

\caption{Higher-vortexable multiorbital zero-mode complexes, illustrated for $n=3$. (a) Constant-$\hat B$ representative $\widehat D_{3,\gamma_0}$ and the exact components of $\bm\Phi_k=(\widehat A_3^\dagger)^k\bm\Phi_0$, with $\hat a^\dagger=a^\dagger+h$. (b) Vortex multiplication induces $\bm\Phi_k\mapsto k\lambda\bm\Phi_{k-1}$, producing one length-four Jordan chain, and a third-vortexable degenerate band complex. (c) A decomposable complex would split into shorter chains; here it decomposes into two first-vortexable chains. The Jordan-chain criterion is invariant under the local band-frame transformations commuting with $z$ considered here.}
\label{fig:chain}
\end{figure}

The remaining zero-mode families are generated from \(\bm{\Phi}_0=(\phi_0,0,\ldots,0)^T\) by a matrix raising operator,
\begin{equation}
A_n^\dagger
=a^\dagger\mathbbm{1}
+\sum_{j=0}^{n-1}\frac{j+1}{n-j}|j+1\rangle\langle j|,
\label{eq:vortexladder}
\end{equation}
through \(\bm{\Phi}_k=(A_n^\dagger)^k\bm{\Phi}_0\), as illustrated for \(n=3\) in Fig.~\ref{fig:chain}(a). Each family is annihilated by \(D_n\) (Supplementary Material). Crucially, since \([z,A_n^\dagger]=\lambda\mathbbm{1}\), multiplication by the vortex coordinate obeys
\[
z\bm{\Phi}_k[\phi_0]
=\bm{\Phi}_k[z\phi_0]
+k\lambda\bm{\Phi}_{k-1}[\phi_0].
\]
The vortex coordinate therefore connects all \(n+1\) families in a single lowering chain, as shown in Fig.~\ref{fig:chain}(b), rather than the shorter chains characteristic of a decomposable complex [Fig.~\ref{fig:chain}(c)]. For nonzero mean flux, the resulting chain is indecomposable under the allowed local band-frame transformations (Supplementary Material).

The construction also allows \(a^\dagger\to\hat a^\dagger=a^\dagger+h(\mathbf r)\), with the corresponding replacement \(B\to\hat B=B+[a,h]\) throughout Eqs.~\eqref{eq:D_n} and \eqref{eq:vortexladder}. Choosing \(\hat B=-\gamma_0\) constant eliminates all longer-range interorbital couplings, leaving nearest-neighbor hoppings \(n\gamma_0,(n-1)\gamma_0,\ldots,\gamma_0\), as illustrated in Fig.~\ref{fig:chain}(a). A constant diagonal similarity transformation makes these hoppings equal without changing the vortexability structure, yielding a uniform-hopping rhombohedral representative that remains first order in momentum. Periodicity requires \(\gamma_0=-\langle B\rangle\); hence a nonzero uniform-hopping representative exists only at nonzero mean flux. At zero mean flux, this representative becomes diagonal and decomposable.

\textit{Discussion.--}
We have designed local Hamiltonians with exactly flat, higher-vortexable Chern bands in a spatially varying magnetic field, generalizing the Aharonov--Casher lowest band. For a scalar particle, a flat first-vortexable band at higher energy requires the kinetic coefficient to be $w\propto B^{-1}$, and this is the only local second-order Hamiltonian that works. The Dirac Hamiltonian has the same structure, with a local velocity $\propto B^{-1/2}$. In chiral Bernal-stacked bilayer graphene with $\braket{B}\neq0$, the zero modes remain exactly degenerate for any $\gamma(\rr)$ and form zeroth- and first-vortexable bands if and only if $\braket{\gamma}\neq0$, so uniform hopping suffices and need not be matched to the field. In the multiorbital generalization, exact zero modes give bands of arbitrary vortexability with a Hamiltonian that is first-order in momentum, including chiral rhombohedral multilayers. 

Beyond Landau-level mimicry, degenerate vortexable bands may qualitatively enrich the fractional quantum Hall problem by providing a multiorbital Hilbert space with distinct form factors.  Bernal-stacked multilayer graphene in a uniform magnetic field already offers a prominent example, supporting competition among Laughlin/Jain, paired and interorbital fractional quantum Hall phases \cite{papic2014topological, zibrov2017tunable, chen2024tunable}. It is therefore natural to ask how this phase structure evolves in the nonuniform magnetic field models considered here.  Related control has already been demonstrated for bosons in optical flux lattices, where engineering the effective field profile can enhance the stability of distinct non-Abelian phases~\cite{sommer2025ideal}. More broadly, the interacting phases of degenerate higher-vortexable complexes, in both uniform and nonuniform fields, provide an interesting direction for future work.

Finally, our results provide a starting point for a systematic classification of local higher-vortexable Hamiltonians with  arbitrary momentum order and orbital number. Within second-order scalar models and first-order models with up to two orbitals per chirality, first vortexability is the maximum, and higher orders require additional orbitals. A full classification should identify the minimal local structures needed for $n^{\mathrm{th}}$ vortexability at fixed differential order and orbital number.

\textit{Acknowledgements -- } 
This work was supported by the EPSRC [grant number EP/V062654/1], a Simons Investigator Award [Grant No. 511029] and a Cambridge International Scholarship provided by the Cambridge Trust. For the purpose of open access, the authors have applied a creative commons attribution (CC BY) licence to any author accepted manuscript version arising.

\bibliography{library.bib}

\section{Supplementary Material}

\subsection{Local AC Hamiltonians}
The factorized form of \(H_w\) in Eq.~\eqref{eq:H_w} need not be assumed. It follows from locality, second differential order, formal self-adjointness, and the existence of the exact Aharonov--Casher zero-mode band. Without loss of generality, we choose the energy origin so that the AC partner band lies at zero energy. Thus, let $H$ be a scalar, local, formally self-adjoint differential operator of order at most two satisfying
\begin{equation}
H\phi_0=0,
\quad
\phi_0\in\ker a.
\end{equation}
Locally, the AC operator may be written as
\begin{equation}
a\sim e^{-K}\partial_{\bar z}e^{K},
\end{equation}
so that its complete local zero-mode family is
$\phi_0=f(z)e^{-K}$, with \(f\) an arbitrary holomorphic function. Conjugating \(H\) by the same local gauge factor,
$
\widetilde H
\equiv
e^K H e^{-K}$,
we therefore have
\[
\widetilde H f(z)=0
\]
for every holomorphic \(f\).

The most general scalar differential operator of order at most two can be written as
\begin{align}
\widetilde H={}&
A_{20}\partial_z^2
+A_{11}\partial_z\partial_{\bar z}
+A_{02}\partial_{\bar z}^2
\nonumber\\
&+
A_{10}\partial_z
+A_{01}\partial_{\bar z}
+A_{00}.
\label{eq:generalL}
\end{align}
Acting on a holomorphic function gives
\[
\widetilde H f(z)
=
A_{20}f''(z)
+
A_{10}f'(z)
+
A_{00}f(z).
\]
At any point, the values of \(f\), \(f'\), and \(f''\) may be given independently by a suitable holomorphic function. Hence, we must have that
$A_{20}=A_{10}=A_{00}=0$, and so every surviving term contains a right factor of \(\partial_{\bar z}\), such that
\[
\widetilde H
=
\widetilde R\,\partial_{\bar z},
\]
with \(\widetilde R\) a local first-order scalar operator. Undoing the conjugation gives
\begin{equation}
H=Ra,
\label{eq:rightfactor}
\end{equation}
for some first-order scalar operator \(R\).

We now impose formal self-adjointness. In the conjugated representation, the second-order part of \(\widetilde H\) has the form
$A_{11}\partial_z\partial_{\bar z}
+
A_{02}\partial_{\bar z}^2$.
The adjoint of the pure \(\partial_{\bar z}^2\) term contains a \(\partial_z^2\) contribution. Since \(\widetilde H\) contains no \(\partial_z^2\) term, formal self-adjointness forces
$A_{02}=0$.
Thus the second-order part of \(\widetilde H\) is purely mixed,
$A_{11}\partial_z\partial_{\bar z}$.

Equivalently, in the original representation, the second-order part of \(H\) can be reproduced by an operator of the form $a^\dagger w(\rr)a$, for some real scalar function $w(\rr)$. We choose $w$ so that $a^\dagger w a$ has exactly the same second-order coefficients as $H$, and define
\begin{equation}
\Delta H\equiv H-a^\dagger w a.
\end{equation}
By construction, $\Delta H$ is formally self-adjoint and of differential order at most one. Moreover,
$\Delta H \phi_0=0$
for every
$\phi_0$,
because both $H$ and $a^\dagger w a$ annihilate the full AC zero-mode family.

Conjugating $\Delta H$ by the same local factor, define
$\widetilde{\Delta H}\equiv e^K \Delta H e^{-K}$.
The most general scalar first-order operator has the form
\begin{equation}
\widetilde{\Delta H}
=
B_{10}\partial_z
+
B_{01}\partial_{\bar z}
+
B_{00}.
\end{equation}
Since $\widetilde{\Delta H}$ annihilates every holomorphic function $f(z)$,
$\widetilde{\Delta H} f=B_{10}f'+B_{00}f=0$.
Because $f$ and $f'$ may be chosen independently at a point, we must have
$B_{10}=B_{00}=0$.
Hence, $\widetilde{\Delta H}
=B_{01}\partial_{\bar z}$,
but formal self-adjointness also eliminates this remaining first-order term. Therefore,
$B_{01}=0$ and $\Delta H=0$.
We conclude that
\begin{equation}
H=a^\dagger w(\rr)a.
\label{eq}
\end{equation}
Thus, the factorized kinetic form is derived: it is the most general scalar, local, formally self-adjoint Hamiltonian of differential order at most two having the full AC zero-mode family as an exact flat band at zero energy. The argument above is local. Its essential input is the complete AC solution family \(\phi_0=f(z)e^{-K}\), rather than any particular finite-dimensional basis of torus zero-modes. Periodic boundary conditions enter only when global constraints are subsequently imposed.

The same local argument also gives a useful first-order corollary. Let \(D\) be a scalar, local, first-order operator satisfying
$D\phi_0=0$
for every
$\phi_0\in\ker a$.
As before, define
$\widetilde D\equiv e^K D e^{-K}$, so that $\widetilde D$ annihilates every holomorphic function $f(z)$. Writing
\begin{equation}
\widetilde D
=C_{10}\partial_z
+C_{01}\partial_{\bar z}
+C_{00},
\end{equation}
we obtain
$\widetilde D f
=C_{10}f'+
C_{00}f=0$.
Following previous arguments, we must have that
$C_{10}=C_{00}=0$, and therefore 
$\widetilde D=C_{01}\partial_{\bar z}$.
Undoing the conjugation yields
\begin{equation}
D=\mu(\rr)a,
\end{equation}
for some scalar function $\mu(\rr)$. Thus, any local scalar first-order operator that annihilates the full AC zero-mode space is necessarily proportional to \(a\).

\subsection{Rigidity from the general first-vortexable state}
The argument presented in Eq.~\eqref{eq:a_dagger_argument} may appear to lack generality in first-vortexable states because we used the representative $a^\dagger\phi_0$. We now allow the full scalar canonical freedom: consider the factorized scalar Hamiltonian $H_w$ (Eq.~\eqref{eq:H_w}) and require $H_w\tilde{\phi}_1=E_1\tilde{\phi}_1$, with $\tilde{\phi}_{1}$ defined in Eq.~\eqref{eq:generalstate}. Because $H_wP_0=0$, define $r=[a,h]$, $k=\frac{w(B + r)}{E_1}-1$.
The eigenvalue equation is then equivalent to
\begin{equation}
 Q_0h\phi_0=a^\dagger(k\phi_0).
 \label{eq:krel}
\end{equation}
Applying $a$ gives
\begin{equation}
 r\phi_0=[a,k]a^\dagger\phi_0+
 \bigl([a^\dagger,[a,k]]+Bk\bigr)\phi_0.
 \label{eq:jeteq}
\end{equation}
The left-hand side is proportional only to $f$, whereas $a^\dagger\phi_0$ contains an independent term proportional to $f'$. Since the identity must hold for arbitrary holomorphic $f$, the coefficient multiplying $f'$ must vanish. Hence $[a,k]=0$, after which Eq.~\eqref{eq:jeteq} gives $r=Bk$. Consequently $B+r=B(1+k)$, and the definition of $k$ yields
\begin{equation}
 (1+k)\left(1-\frac{wB}{E_1}\right)=0.
\end{equation}
If $k=-1$, then $B+r=0$, so $(a^\dagger+h)\phi_0\in\ker a$ for every $f$ and the $Q_0$ projection annihilates the entire putative first band. Hence, for a nontrivial band, $k\neq-1$ and we obtain the solution
\begin{equation}
w(\rr)=E_1 B^{-1}(\rr).
\label{eq:wunique}
\end{equation}
Furthermore, using the eigenvalue equation in ~\eqref{eq:krel}, we get that the first-vortexable states are given by
\begin{equation}
 \tilde{\phi}_1=a^\dagger[(1+k)\phi_0],\quad [a,k]=0,
 \label{eq:reparam}
\end{equation}
so the apparently more general first-vortexable eigenspace is only a holomorphic reparametrization of $a^\dagger\ker a$. On a connected magnetic torus, periodic holomorphic $k$ is constant; for periodic $h$ and nonzero mean flux, integrating $[a,h]=Bk$ fixes $k=0$. Thus the first-vortexable eigenspace is recovered exactly as $a^\dagger\ker a$. 
This establishes the first rigidity statement: within the $H_w$ family of Hamiltonians, the function $w(\rr)$ is selected by the \emph{general} first-vortexability condition.

\subsection{Position-dependent-mass interpretation}

Expanding the factorized Hamiltonian $H_w$ in Eq.~\eqref{eq:H_w} without commuting $w$ through the covariant momenta gives
\begin{equation}
 H_w
 =\frac12\bm\Pi\cdot w\bm\Pi
 -\frac{1}{2}wB
 +\frac{1}{2}(\nabla w\times\bm\Pi)_z.
 \label{eq:Hw-expanded}
\end{equation}
The first term has the standard position-dependent-mass form upon identifying $
 w(\rr)=1/M_{\rm eff}(\rr)$.
In particular, the first-vortexability condition $w=E_1 B^{-1}(\rr)$ implies $M_{\rm eff}(\rr)=MB(\rr)$. Thus, first vortexability requires the local effective mass to track the magnetic texture. The first term in Eq.~\eqref{eq:Hw-expanded} is the gauge-covariant BenDaniel--Duke kinetic energy~\cite{bendaniel1966space}, the second is the magnetic-moment term associated with the chosen Pauli block, and the third is an orbital coupling generated by gradients of the effective mass. The orbital term can be absorbed into a shifted vector potential,
\begin{equation}
 H_w=\tfrac12(\bm\Pi-\delta\bm A)\cdot w(\bm\Pi-\delta\bm A)-\tfrac12 wB-\frac{|\nabla w|^2}{8w},
 \label{eq:Hw-minimal}
\end{equation}
with $\delta\bm A=-\tfrac12\hat{\bm z}\times\nabla\ln w$, as can be checked directly. For $w=E_1B^{-1}$, $H^\star_{\rm AC}$ is therefore a standard position-dependent-mass Hamiltonian with $M_{\rm eff}\propto B$ in the field $B+\tfrac12\nabla^2\ln B$ and with scalar potential $-E_1|\nabla B|^2/(8B^3)$, up to a constant.

Due to ordering ambiguities in position-dependent effective mass Hamiltonians~\cite{quesne2005hamiltonians, bravo2016position}, it is useful to compare Eq.~\eqref{eq:Hw-expanded} with the gauge-covariant von Roos family~\cite{von1983position},
\begin{equation}
 T_{\alpha\beta\gamma}
 =
 \frac14\left[
 M^\alpha\bm\Pi M^\beta\bm\Pi M^\gamma
 +
 M^\gamma\bm\Pi M^\beta\bm\Pi M^\alpha
 \right],
\end{equation}
with $\alpha+\beta+\gamma=-1$.
Writing $w=1/M$, this may be expressed as
\begin{equation}
 T_{\alpha\beta\gamma}[w]
 =
 \frac12\bm\Pi\cdot w\bm\Pi
 +
 U_{\alpha\beta\gamma}[w],
\end{equation}
where
\begin{equation}
 U_{\alpha\beta\gamma}[w]
 =
 \frac{(\alpha+\gamma)}{4}\nabla^2w
 +
 \frac{\alpha\gamma}{2}
 \frac{|\nabla w|^2}{w}.
\end{equation}
Accordingly, the same factorized Hamiltonian can be written for any admissible von Roos ordering as
\begin{equation}
 H_w
 =
 T_{\alpha\beta\gamma}[w]
 +
 V_{\alpha\beta\gamma} [w],
\end{equation}
with
\begin{equation}
 \begin{split}
 V_{\alpha\beta\gamma}[w]
 ={}&
 -U_{\alpha\beta\gamma}[w]
 -\frac{1}{2}wB
 +\frac{1}{2}(\nabla w\times\bm\Pi)_z .
 \end{split}
\end{equation}
Thus, changing the von Roos ordering only repartitions the fixed operator $a^\dagger w a$ between the kinetic term and compensating counterterms. The BenDaniel--Duke choice $(\alpha,\beta,\gamma)=(0,-1,0)$ is the simplest representation because it introduces no additional scalar ordering terms; vortexability selects the full factorized Hamiltonian rather than a particular ordering convention.

\subsection{Comparison with uniform magnetic field on a curved surface}

Consider the conformal metric
$g_{ij}=\sqrt{g(\rr)}\delta_{ij}$.
The magnetic-field two-form is $F=B(\bm r)\,dx\wedge dy$, and the
scalar magnetic field measured in the curved geometry is $B_g=F_{xy}/\sqrt g=B(\bm r)/\sqrt{g}$.
Thus, the choice
$\sqrt{g}(\rr)=B(\rr)$
makes \(B_g=B_0\) spatially uniform.  This construction requires
$B>0$; a zero or sign change of $B$ would make the metric degenerate
or indefinite. The scalar Hilbert space on the curved surface is now
\begin{equation}
 \mathcal H_g=L^2(\R^2,\Omega^2d^2r), \quad
 \ipcurved{\psi}{\chi}
 =\int d^2r\,\Omega^2\psi^*\chi.
\end{equation}
The unitary map from curved to flat measure is
\begin{equation}
 U:\mathcal H_g\longrightarrow\mathcal H_0,
 \quad
 U\psi=\Omega\psi.
 \label{eq:unitaryU}
\end{equation}
For the spin-polarised scalar Pauli block, the standard covariant kinetic
operator is \cite{lawson2016spin, estienne2023ideal}
\begin{equation}
 H_{g}
 =
 \frac{1}{2M}
 \left[
 \frac{1}{\sqrt g}\Pi_i\!\left(\sqrt g\,g^{ij}\Pi_j\right)
 -B_g
 \right].
 \end{equation}
In two conformal dimensions, $\sqrt g\,g^{ij}=\delta^{ij},$
so
\begin{equation}
\begin{aligned}
 H_{g}
 &=
 \frac{1}{2M}g^{-1/2}
 \left(\Pi_x^2+\Pi_y^2- B (\rr)\right)\\
 &=
 \frac{1}{M}g^{-1/2}a^\dagger a.
 \label{eq:Hgcurvedmeasure}
\end{aligned}
\end{equation}
It is important to note that this left-ordered expression is hermitian only in $\mathcal H_g$, not in
$\mathcal H_0$. Transporting it to the flat measure gives
\begin{equation}
\begin{aligned}
 H_{g\to0}
 &=
 UH_{g}U^{-1}\\
 &=
 \frac{1}{M}g^{-1/4}a^\dagger ag^{-1/4}.
 \label{eq:Hgflatmeasure}
\end{aligned}
\end{equation}
Equivalently, $ H_{g\to0}\propto\,(ag^{-1/4})^\dagger(ag^{-1/4})$. Using $\sqrt{g}=B(\rr)$, we compare the first-vortexable Hamiltonian $H^{\star}_{\mathrm{AC}}$
\begin{equation}
H^{\star}_{\mathrm{AC}}
\propto
\left(B^{-1/2}(\rr)\,a\right)^\dagger
\left(B^{-1/2}(\rr)\,a\right)
\end{equation}
with
\begin{equation}
H_{g\to0}\propto
\left(a B^{-1/2}(\rr)\right)^\dagger
\left(aB^{-1/2}(\rr)\right).
\end{equation}
Therefore, $H^{\star}_{\mathrm{AC}}\neq H_{g\to0}$ for nonconstant $B(\rr)$, due to a subtle ordering distinction. 

\subsection{Symmetric scalar dressing}

The preceding curved-space comparison motivates enlarging the scalar Hamiltonian class beyond the simple factorization $a^\dagger w a$. A natural Hermitian generalization is obtained by allowing a real multiplicative dressing on both sides,
\begin{equation}
    H_{h,w}
    =
    h(\rr)a^\dagger w(\rr)a h(\rr),
    \label{eq:hwh_H}
\end{equation}
where $h(\rr)$ and $w(\rr)$ are smooth real functions and $h$ is nowhere vanishing. Unlike a change of $w$ alone, a nonconstant $h$ also changes the natural zero-mode space,
\begin{equation}
    \ker H_{h,w}=h^{-1}\ker a,
    \label{eq:hwh_zero_modes}
\end{equation}
under the same positivity assumptions used above. Thus this class allows a genuine deformation of the lowest-vortexable parent rather than merely a different ordering of the same AC zero modes.

We now ask whether this enlarged scalar class admits additional exact first-vortexable parents. Defining $\tilde{a} \equiv ah$,
Eq.~\eqref{eq:hwh_H} becomes $H_{h,w}=\tilde{a}^\dagger w \tilde{a}$. A generic zero mode may be written as $\psi_0=h^{-1}\phi_0$. The corresponding natural first-vortexable candidate is
\begin{equation}
\tilde{a}^\dagger\psi_0
= h a^\dagger h^{-1}\phi_0
= \bigl(a^\dagger+\eta_h\bigr)\phi_0,
\label{eq:hwh_first_state}
\end{equation}
where $\eta_h\equiv h[a^\dagger,h^{-1}]$. Suppose this state forms a nontrivial flat band at energy $E_1\neq0$. Since $\tilde{a}\psi_0=0$,
\begin{equation}
    \tilde{a}^\dagger w[\tilde{a},\tilde{a}^\dagger]\psi_0
    =
    E_1 \tilde{a}^\dagger\psi_0.
\end{equation}
Because $h$ is nowhere vanishing, $\tilde{a}^\dagger$ is injective whenever $a^\dagger$ is, and therefore
\begin{equation}
w[\tilde{a},\tilde{a}^\dagger]\psi_0
= E_1\psi_0.
\label{eq:hwh_commutator_eigen}
\end{equation}
The commutator is
\begin{equation}
    [\tilde{a},\tilde{a}^\dagger]
    =
    h^2B
    +h[a,h]a^\dagger
    -h[a^\dagger,h]a
    -h[a^\dagger,[a,h]].
    \label{eq:A_commutator_expanded}
\end{equation}
When Eq.~\eqref{eq:A_commutator_expanded} acts on
$h^{-1}f(z)\mathcal N_0$, the term
$h[a,h]a^\dagger$ contains an independent contribution proportional to $f'(z)$, whereas the right-hand side of Eq.~\eqref{eq:hwh_commutator_eigen} contains only $f(z)$. Since the relation must hold for arbitrary holomorphic $f$, the coefficient of this independent derivative must vanish, $[a,h]=0$.
For real $h$, this implies that $h$ is constant and may be absorbed into $w$. The enlarged Hamiltonian class therefore collapses to the scalar class considered previously, for which first vortexability requires
$w \propto B^{-1}$. Thus, a real symmetric multiplicative dressing produces no new scalar first-vortexable parent.

The restriction to real $h$ is essential. For a complex nonvanishing dressing, the Hermitian factorization is
$(ah)^\dagger w(ah)=h^*a^\dagger w a h$. A nonconstant holomorphic factor can then represent a change of holomorphic frame of the vortexable zero-mode bundle rather than a physically distinct kinetic dressing. The above result concerns the real positive dressings relevant to curved-space interpretations.

\subsection{General nonuniform Bernal block and first vortexability}
Consider the off-diagonal differential operator of Bernal graphene
\begin{equation}
 D_{1,\gamma}=
 \begin{pmatrix}a&\gamma(\rr)\\0&a\end{pmatrix},
\end{equation}
where both the magnetic field ($[a,\adag]=B(\rr)$) and interlayer tunneling $\gamma(\rr)$ are allowed to be spatially varying. One zero-mode family is $\bm\phi_0 =(\phi_0,0)^T$. For the second, $\bm{\phi}_1 = (\psi_1, \psi_2)$ and $\psi_2=\phi_0\in\ker a$, which gives the following differential equation for $\psi_1$:
\begin{equation}
 a\psi_1=-\gamma\phi_0.
\end{equation}
Explicitly, in the gauge $A=2\ii\partial_{\bar z}K$ one has $a=-\sqrt2\,\ii\,\ee^{-K}\partial_{\bar z}\ee^{K}$, so a particular solution is $\psi_1=G(\rr)f(z)\ee^{-K}$ with $\partial_{\bar z}G=-\ii\gamma/\sqrt2$, generalizing the polyanalytic zero modes found in Ref.~\cite{kailasvuori2009pedestrian} for the effective quadratic-in-momentum model.
We do not assume a first-vortexable solution, but instead ask for a local first-order solution operator valid for arbitrary AC zero modes. Modulo addition of some $\chi_0\in\ker a$, its most general action on $\ker a$ can be written as the following ansatz:
\begin{equation}\psi_1=(k(\rr)a^\dagger+h(\rr))\phi_0+\chi_0.
\end{equation}
Then
\begin{equation}
 a\psi_1=[a,k]a^\dagger\phi+(kB+[a,h])\phi.
\end{equation}
The right-hand side of the zero-mode equation contains only the local holomorphic function $f$, whilst $a^\dagger\phi_0$ contains an independent $f'$ term. Therefore,
\begin{equation}
[a,k]=0,\quad \gamma=-kB-[a,h].
\label{eq:gamma_constraint} 
\end{equation}
Now use $[a,z]=0$ and define $\lambda\equiv[z,a^\dagger]$, which is constant. Then
\begin{equation}
 z\binom{(ka^\dagger+h)\phi}{\phi}
 =
 \binom{(ka^\dagger+h)(z\phi)}{z\phi}
 +\binom{\lambda k\phi}{0}.
\end{equation}
Because $[a,k]=0$, $k\phi\in\ker a$, multiplication by $z$ closes on the two zero-mode families. On a connected torus a periodic holomorphic $k$ is constant. If $\braket B\neq0$, integrating Eq.~\eqref{eq:gamma_constraint} gives $
k=-\braket\gamma/\braket B$.
Hence $\braket\gamma\neq0$ fixes a nonzero lowering link. Conversely, if $\braket\gamma=0$, the periodic equation $\gamma=-[a,h_0]$ has a solution, and $(h_0\phi_0,\phi_0)^T$ is separately closed under $z$. The two-band complex is then decomposable. Thus, for nonzero mean flux the periodic Bernal complex is genuinely first-vortexable precisely when $\braket\gamma\neq0$.

If $\braket B=0$ and $\braket\gamma\neq0$, the average of Eq.~\eqref{eq:gamma_constraint} is inconsistent. If $\braket B = \braket \gamma =0$, the periodic equation $\gamma=-[a,h_0]$ admits a solution. Consequently, the two zero-mode families $(\phi_0,0)^T$ and $(h_0\phi_0,\phi_0)^T$ are separately closed under multiplication by $z$, so the AC zero-mode complex is decomposable. Although the averaged constraint does not fix $k$ in this case, a nonzero $k$ does not imply an intrinsic lowering link of the vortexable complex. Indeed, since $\braket{B}=0$, there exists a periodic $h_{B}$ satisfying $[a,h_{B}]=-B$, and hence $(a^\dagger+ h_{B})\phi_0\in\ker a$. Therefore, any solution with nonzero $k$ can be written as $(ka^\dagger+h)\phi_0=h_0\phi_0+k(a^\dagger+t)\phi_0$, differing from the decomposed family only by an element of the lower AC sector. Thus, any apparent lowering link is removable by a change of zero-mode basis. In particular, the smooth periodic Bernal AC construction does not support a genuinely indecomposable first-vortexable two-band complex at zero mean flux.

When $k\neq0$, rescaling the second family gives the canonical form
\begin{equation}
\bm\phi_1=
\binom{Q_0(a^\dagger+h)\phi_0}{c\phi_0},
\qquad c=1/k,
\label{eq:firstvortexableBernal}
\end{equation}
with a correspondingly rescaled $h$. Its zero-mode condition is $\gamma=-(B+[a,h])/c$, and the representative $h=0$ has $\gamma\propto B$.

\subsubsection{Absence of a second-vortexable complex}

We now show by contradiction that the two-orbital Bernal Hamiltonian \eqref{eq:Bernal} cannot support a genuine second-vortexable band. 
Suppose first that the full Hamiltonian possessed a genuine second-vortexable eigenstate
\begin{equation}
H_{\rm B}
\begin{pmatrix}
\bm \phi_2\\
\bm \psi_2
\end{pmatrix}
=E_2\begin{pmatrix}
\bm \phi_2\\
\bm \psi_2
\end{pmatrix}.
\label{eq:Bernal_second_eigen}
\end{equation}
Since the chiral projection commutes with multiplication by $z$, the upper component $\bm \phi_2$ must close under vortex multiplication into the first-vortexable zero mode in Eq.~\eqref{eq:firstvortexableBernal}. Within the local second-vortexable class, the most general form needed for this closure is
\begin{equation}
\bm \phi_2[\phi_0]
=
\begin{pmatrix}
(F(a^\dagger)^2+p\,a^\dagger+q)\phi_0\\
(s\,a^\dagger+t)\phi_0
\end{pmatrix},
\label{eq:Bernal_u2}
\end{equation}
where $F,p,q,s,t$ are position-dependent  operators. Genuine second vortexability requires $F\neq0$.

The second-vortexability closure condition may be written as
\begin{equation}
z\bm \phi_2[\phi_0]-\bm \phi_2[z\phi_0]
=
\lambda\,\bm \phi_1[s\phi_{0}].
\label{eq:Bernal_second_closure}
\end{equation}
For the state on the right-hand side to belong to the first-vortexable band for arbitrary $\phi_0\in\ker a$, one must have
$[a,s]=0$.
Comparing the two components of Eq.~\eqref{eq:Bernal_second_closure} then gives
\begin{equation}
2F=s/c,
\quad
p=\frac{[a^\dagger,s]+hs}{c}.
\label{eq:2F_constraint}
\end{equation}
Since $c$ is constant, this implies
$[a,F]=0$ i.e. that $F$ is holomorphic.
Writing $\eta\equiv[a,h]$,
and using $[a,s]=0$ together with the Jacobi identity gives
\begin{equation}
[a,p]
=
\frac{s}{c}\eta.
\label{eq:[a,p]}
\end{equation}
We now act with $D_1$ on Eq.~\eqref{eq:Bernal_u2}. Using
\begin{equation}
a(a^\dagger)^2\phi_0
=
2Ba^\dagger\phi_0
+
[a^\dagger,B]\phi_0,
\end{equation}
the coefficient of $a^\dagger\phi_0$ in the first component is
\begin{equation}
2FB+[a,p]+\gamma s.
\label{eq:upper_comp_coeff_adagger}
\end{equation}
Equations~\eqref{eq:gamma_constraint}, \eqref{eq:2F_constraint}, \eqref{eq:[a,p]}, and \eqref{eq:upper_comp_coeff_adagger} give
\begin{equation}
\begin{aligned}
2FB+[a,p]+\gamma s
&=
2FB
+\frac{s}{c}\eta
-\frac{s}{c}(B+\eta)
\\
&=
\left(2F-\frac{s}{c}\right)B
=0.
\end{aligned}
\end{equation}
The $a^\dagger\phi_0$ contribution therefore cancels identically and one obtains
\begin{equation}
D_1\bm \phi_2[\phi_0]
=
\begin{pmatrix}
M_1(\rr)\phi_0,
M_2(\rr)\phi_0
\end{pmatrix}^{T},
\label{eq}
\end{equation}
where
\begin{equation}
M_1
=
F[a^\dagger,B]
+pB+[a,q]+\gamma t,
\quad
M_2
=
sB+[a,t].
\label{eq}
\end{equation}
Thus $D_1$ maps the assumed second-vortexable state entirely into the lowest AC band: no $a^\dagger\phi_0$ or $(a^\dagger)^2\phi_0$ term survives.

From this analysis, it follows that
$D_1^\dagger D_1\bm \phi_2$
contains at most one power of $a^\dagger$ acting on $\phi_0$. On the other hand, squaring the full chiral eigenvalue equation gives
\begin{equation}
D_1^\dagger D_1\bm \phi_2
=
E_2^2\bm \phi_2.
\label{eq:second_order_phi_2_eq}
\end{equation}
The first component on the right-hand side contains the term $E_2^2F(a^\dagger)^2\phi_0$.
Since the identity must hold for arbitrary holomorphic $f$ in $\phi_0$, and this second-derivative contribution is independent of the lower holomorphic derivatives, Eq.~\eqref{eq:second_order_phi_2_eq} requires
\begin{equation}
E_2^2F=0.
\end{equation}
For a genuine second-vortexable state, $F\neq0$ and therefore
$E_2=0$.
This implies that $D_1\bm \phi_{1}=0$, but we already know that all zero modes of 
$D_1$ may be written as
\begin{equation}
\bm \phi[\chi_0,\phi_0]
=
\begin{pmatrix}
\chi_0+(a^\dagger+h)\phi_0\
, c\phi_0
\end{pmatrix}^{T},
\quad
\chi_0,\phi_0\in\ker a,
\label{eq}
\end{equation}
so the zero-energy sector contains only the lowest-vortexable band and its first-vortexable partner. It cannot supply the independent third band required for a genuine second-vortexable complex. (Note that the lower component must always be proportional to an AC zero-mode, which would imply $s=F=0$ in Eq.~\eqref{eq:Bernal_u2}.)
Hence, the two-orbital Bernal Hamiltonian admits no genuine second-vortexable eigenband, at zero or nonzero energy.
The obstruction is independent of the spatial profile of $\gamma(\rr)$ and follows from the two-orbital first-order structure together with the closure condition of canonical second vortexability.

\subsection{Arbitrary-order first-order multiorbital construction}

The higher-vortexability hierarchy can be constructed explicitly from the nonuniform commutator $[a,a^\dagger]=B(\rr)$.
For any integer $n\geq1$, the exact identity
\begin{equation}
    [a,(a^\dagger)^n]
    =
    \sum_{\ell=0}^{n-1}
    (a^\dagger)^\ell
    B
    (a^\dagger)^{n-1-\ell}
\end{equation}
may be brought into normal-ordered form using
\begin{equation}
    (a^\dagger)^\ell B
    =
    \sum_{s=0}^{\ell}
    {\ell\choose s}
    \operatorname{ad}_{a^\dagger}^{(s)}(B)
    (a^\dagger)^{\ell-s},
\end{equation}
where $\operatorname{ad}^{(s+1)}_{a^\dagger}(X) = [a^\dagger,\operatorname{ad}^{(s)}_{a^\dagger}(X)]$ ($\operatorname{ad}^{(0)}_{a^\dagger}(X)\equiv X$).
Summing the resulting binomial coefficients with the hockey-stick identity gives, for any $\phi_0\in\ker a$,
\begin{equation}
a(a^\dagger)^k\phi_0
=\sum_{r=0}^{k-1}{k\choose r}
\operatorname{ad}_{a^\dagger}^{(k-1-r)}(B)
(a^\dagger)^r\phi_0.
\label{eq:higher_commutator_action}
\end{equation}
Since $a^\dagger$ is first-order and $B$ is a multiplication operator, every nested commutator
$\operatorname{ad}_{a^\dagger}^{(s)}(B)$ is again multiplication by a local derivative of the magnetic texture.

Consider now the $(n+1)$-component state
\begin{equation}
    \mathbf{\Phi}_{n}
    =
    \bigl(
    (a^\dagger)^n\phi_0,\,
    (a^\dagger)^{n-1}\phi_0,\,
    \ldots,\,
    a^\dagger\phi_0,\,
    \phi_0
    \bigr)^T .
    \label{eq:Phi_n_general}
\end{equation}
Indexing rows and columns by $i,j=0,\ldots,n$, define the upper-triangular first-order operator
\begin{equation}
    (D_n)_{ij}
    =
    \begin{cases}
    a,
    & i=j,\\[3pt]
    -\displaystyle {n-i\choose n-j}\,
    \operatorname{ad}_{a^\dagger}^{(j-i-1)}(B),
    & i<j,\\[6pt]
    0,
    & i>j.
    \end{cases}
    \label{eq:Dn_general}
\end{equation}
Equation~\eqref{eq:higher_commutator_action} then ensures, row by row, that $D_n\mathbf{\Phi}_n=\mathbf{0}$.
The first few members are
\begin{equation}
    D_1=
    \begin{pmatrix}
        a&-B\\
        0&a
    \end{pmatrix},
\end{equation}
\begin{equation}
    D_2=
    \begin{pmatrix}
        a&-2B&-[a^\dagger,B]\\
        0&a&-B\\
        0&0&a
    \end{pmatrix},
\end{equation}
and
\begin{equation}
    D_3=
    \begin{pmatrix}
        a&-3B
        &-3[a^\dagger,B]
        &-[a^\dagger,[a^\dagger,B]]
        \\
        0&a&-2B&-[a^\dagger,B]
        \\
        0&0&a&-B
        \\
        0&0&0&a
    \end{pmatrix}.
\end{equation}
For a uniform field, all nested commutators of $B$ vanish and $D_n$ reduces to the familiar bidiagonal Landau-level/Bernal structure (where $B=1$). For a nonuniform field, the additional matrix elements supply precisely the local derivative corrections required to cancel the obstruction generated by commuting $a$ through successive powers of $a^\dagger$. 

\subsubsection{Full $n$-vortexable complex of zero modes}

Consider the $(n+1)$-component states
\begin{equation}
\mathbf{\Phi}_k[\phi_0]
=
\left(A_n^\dagger\right)^k
\mathbf{\Phi}_0[\phi_0],
\quad
k=0,\ldots,n,
\end{equation}
where
$\mathbf{\Phi}_0[\phi_0]
=\begin{pmatrix}\phi_0, 0, \cdots, 0
\end{pmatrix}^{T}$,
and $A_n^\dagger$ is chosen such that
$[z,A_n^\dagger]=\lambda\,\mathbbm 1$.
It follows immediately that
\begin{equation}
[z,(A_n^\dagger)^k]
=k\lambda( A_n^\dagger)^{k-1},
\end{equation}
and hence we get one property of vortexability satisfied (indecomposability is discussed in the following subsection):
\begin{equation}
z\mathbf{\Phi}_k[\phi_0]
=
\mathbf{\Phi}_k[z\phi_0] + k\lambda\mathbf{\Phi}_{k-1}[\phi_0].
\label{eq:vortex_lowering}
\end{equation}
In addition, requiring $(A_{n}^{\dagger})^{n}\mathbf{\Phi}_{0} = \mathbf{\Phi}_{n}$, where $\mathbf{\Phi}_{n}$ is defined in Eq.~\eqref{eq:Phi_n_general}, we get that the required operator $A^{\dagger}_{n}$ takes the form
\begin{equation}
A_n^\dagger
= a^\dagger\mathbbm{1}+
\sum_{j=0}^{n-1}\frac{j+1}{n-j}|j+1\rangle\langle j|.
\label{eq:vortexladder_SM}
\end{equation}

We now show that $\mathbf{\Phi}_{k}$ are zero-modes of $D_n$, for $k=0,\ldots,n$. Write
$D_n=a\mathbbm{1}+U_n$, $A_n^\dagger
= a^\dagger\mathbbm{1}+S_n$,
with
\begin{equation}
U_n
=
-\sum_{i<j}\binom{n-i}{n-j}\operatorname{ad}_{a^\dagger}^{(j-i-1)}(B)\ket{i}\bra{j}
\end{equation}
and
\begin{equation}
S_n
=
\sum_{j=0}^{n-1}
\frac{j+1}{n-j}|j+1\rangle\langle j|.
\end{equation}
Since $S_n$ does not contain functions of position,
\begin{equation}
[D_n,A_n^\dagger]
=
B \mathbbm{1}
+[U_n,a^\dagger]
+[U_n,S_n].
\end{equation}
Remarkably, we find that all matrix elements cancel except those in the final column, thereby obtaining:
\begin{equation}
[D_n,A_n^\dagger]
=
\sum_{i=0}^{n}
\frac{n+1}{n-i+1}
\operatorname{ad}_{a^\dagger}^{(n-i)}(B)
|i\rangle\langle n|.
\label{eq:Dn_An_commutator}
\end{equation}
Thus, the failure of $D_n$ and $A_n^\dagger$ to commute is entirely confined to the last orbital component.

Since $
\mathbf\Phi_k =( A_n^\dagger)^k\mathbf\Phi_0$
has support only on the first $k+1$ orbital components, its last component vanishes for $k<n$. Hence,
\begin{equation}
[D_n,A_n^\dagger]\mathbf\Phi_k=0,
\quad k<n.
\end{equation}
Starting from $D_n\mathbf\Phi_0=0$, it follows recursively that
\begin{equation}
D_n\mathbf\Phi_{k+1}
=A_n^\dagger D_n\mathbf\Phi_k
+ [D_n, A_n^\dagger]\mathbf\Phi_k
=0,
\end{equation}
and therefore
\begin{equation}
D_n\mathbf\Phi_k=0,
\quad k=0,\ldots,n.
\end{equation}
Thus $ A_n^\dagger$ generates the complete zero-mode hierarchy while preserving the kernel of $D_n$ up to the top of the $(n+1)$-orbital ladder.

\subsubsection{Indecomposability of $n$-vortexable complex}

Equation~\eqref{eq:vortex_lowering} defines a lowering map $\mathsf L$ on the band complex,
\begin{equation}
\mathsf L\mathbf{\Phi}_k
\equiv
z\mathbf{\Phi}_k-\mathbf{\Phi}_k z
=
k\lambda\mathbf{\Phi}_{k-1},
\quad
\mathsf L\mathbf{\Phi}_0=0,
\label{eq:L_definition}
\end{equation}
where the second occurrence of $z$ denotes multiplication by $z$ on the underlying state $\phi_0$. In the frame of ordered vortexable bands
$\{\mathbf{\Phi}_0,\mathbf{\Phi}_1,\ldots,\mathbf{\Phi}_n\}$, $\mathsf L$ can be written as the matrix
\begin{equation}
\mathsf L
=
\lambda
\begin{pmatrix}
0 & 1 & 0 & \cdots & 0\\
0 & 0 & 2 & \cdots & 0\\
0 & 0 & 0 & \ddots & \vdots\\
\vdots & \vdots & \ddots & \ddots & n\\
0 & 0 & \cdots & 0 & 0
\end{pmatrix}.
\label{eq:L_matrix}
\end{equation}
Upon the rescaling $\mathbf{\Phi}_k\rightarrow\frac{1}{k!}\mathbf{\Phi}_k$, $\mathsf L$ is proportional to a nilpotent Jordan matrix of dimension $n+1$. In particular,
\begin{equation}
\mathsf L^{n+1}=0,
\quad
\mathsf L^n\neq0,
\label{eq:nilpotency_index}
\end{equation}
since $
\mathsf L^n\Phi_n
=
n!\lambda^n\Phi_0\neq0$.
Thus, the nilpotency index of the vortexability-lowering map is exactly $n+1$.

The nilpotency of $\mathsf L$ establishes the indecomposability of the vortexability chain. To see this, suppose that an admissible invertible change of band frame decomposed the $(n+1)$-band complex into independent vortexability chains of lengths
$\ell_1,\ldots,\ell_m$, $\sum_{\alpha=1}^{m}\ell_\alpha=n+1$,
with every chain shorter than the original one, $\ell_\alpha<n+1$.
In a frame adapted to this decomposition, the lowering map would have the Jordan form
\begin{equation}
\mathsf L'
 =  J_{\ell_1}\oplus\cdots\oplus J_{\ell_m},
\end{equation}
and therefore $
(\mathsf L')^{\ell_{\rm max}}=0$ (
$\ell_{\rm max}\equiv\max_\alpha\ell_\alpha\le n$).
In particular,
\begin{equation}
(\mathsf L')^n=0.
\label{eq:decomposed_nilpotency}
\end{equation}
For an admissible band-frame transformation $S$ satisfying $[z,S]=0$, however,
$\mathsf L'=S^{-1}\mathsf L S$,
so that
\begin{equation}
(\mathsf L')^n
=
S^{-1}\mathsf L^nS.
\end{equation}
Equation~\eqref{eq:decomposed_nilpotency} would therefore imply
$\mathsf L^n=0$, in contradiction with
Eq.~\eqref{eq:nilpotency_index}. Hence, the complex cannot be decomposed into a direct sum of shorter vortexability chains and the states
$\{\mathbf{\Phi}_0,\ldots,\mathbf{\Phi}_n\}$ form a single genuine $(n+1)$-level vortexability chain.

The indecomposability argument also extends to local differential changes of zero-mode frame, which need not commute with $z$. Under such transformations, the lowering map obeys
\begin{equation}
\mathsf L'=S^{-1}\mathsf LS+S^{-1}[z,S].
\end{equation}
The crucial distinction is whether the mean magnetic flux vanishes. For $\langle B\rangle\neq0$, any smooth, periodic, finite-order differential operator preserving arbitrary local AC zero modes must act as a constant on the AC space. This follows by expanding the operator in powers of $a^\dagger$: the highest derivative coefficient is constant, while averaging the equation for the next coefficient forces it to vanish when $\langle B\rangle\neq0$. Repeating this argument eliminates all derivatives. Thus, any admissible change of frame acts as a constant matrix on the AC space, so that $[z,S]=0$ there. The lowering map consequently retains its single Jordan block, and the vortexability chain cannot be decomposed.

At zero mean flux, by contrast, the chain can be explicitly decomposed. Indeed, $\langle B\rangle=0$ guarantees a periodic function $h_{B}$ satisfying $[a,h_{B}]=-B$. The operator $P=a^\dagger+h_{B}$ then preserves the scalar AC space and satisfies $[z,P]=\lambda$. Since $\mathsf L$ is nilpotent, the local differential transformation
$S=\exp(-P\mathsf L/\lambda)$ is a finite polynomial and gives $\mathsf L'=0$, completely decoupling the vortexable families. Thus, nonzero mean flux protects the indecomposable vortexability chain, while at zero mean flux it can be unwound into independent lowest-vortexable sectors.

\subsubsection{Generalisation of $D_n$}

It is possible to generalise the class of higher-vortexable Hamiltonians $D_{n}$. The simplest extension involves the replacement $\adag\rightarrow \hat{a}^{\dagger} = \adag + h(\rr)$, motivated by the general first-vortexable solution in Eq.~\eqref{eq:generalstate}. To annihilate the corresponding $n+1$-component state $\widehat{\bm\Phi}_{n}=\bigl((\hat{a}^\dagger)^n\phi_0,\,
(\hat{a}^\dagger)^{n-1}\phi_0,\,
\ldots,\,\hat{a}^\dagger\phi_0,\,\phi_0\bigr)^T$, one can show that the same construction as in Eq.~\eqref{eq:Dn_general} works to annihilate the state, but with the replacement $B\rightarrow B + [a,h]$. In fact, the whole construction is preserved with every instance of $\adag$ replaced with $\hat{a}^{\dagger}$ in $A_{n}^{\dagger}$, and guarantees $\widehat{D}_{n} \widehat{\Phi}_{k} = 0$ for $k=0,\ldots,n$. Note that this generalisation now includes the full class of Hamiltonians $D_{1,\gamma}$ for $n=1$.

An important corollary is that we can again choose $[a,h]$ such that $[a,h]+B=-\gamma_0$ is a constant, and all position-dependent elements of $\widehat{D}_{n}$ vanish. As in the Bernal case, we must again assume $\braket{B}\neq 0$ (otherwise $\gamma_0=0$). With this, we have shown the $n$-th vortexability of the operator
\begin{equation}
\widehat{D}_{n,\gamma_0}
=
\begin{pmatrix}
a & n\gamma_0 & 0 & \cdots & 0\\
0 & a & (n-1)\gamma_0 & \cdots & 0\\
0 & 0 & a & \ddots & \vdots\\
\vdots & \vdots & \ddots & \ddots & \gamma_0\\
0 & 0 & \cdots & 0 & a
\end{pmatrix},
\end{equation}
where we stress again that $a$ is defined with respect to a non-uniform magnetic field $B(\rr)$. A constant diagonal rescaling $S$ of the orbitals, $\widehat{D}_{n,\gamma_0}\to S\widehat{D}_{n,\gamma_0}S^{-1}$, makes all superdiagonal entries equal. Since $S$ commutes with $z$, the zero modes $S\widehat{\bm\Phi}_k$ keep the same chain structure, and the rescaled operator is the chiral rhombohedral $(n+1)$-layer graphene operator with uniform hopping.

\subsubsection{Higher vortexability and increasing momentum order}
For comparison, consider the familiar higher-order chiral Dirac family
\begin{equation}
H_{\mathrm D}^{(n)}
=
\Delta
\begin{pmatrix}
0 & (a^\dagger)^n\\
a^n & 0
\end{pmatrix},
\label{eq:mudiracfamily_SM}
\end{equation}
whose zero modes in the relevant chiral sector span $\ker a^n$. The upper components of our multiorbital zero modes may be written as
\begin{equation}
\phi_k=(a^\dagger+h)^k\phi_0,
\qquad
k=0,\ldots,n-1,
\end{equation}
with $a\phi_0=0$. Using Eq.~\eqref{eq:higher_commutator_action} with $a^\dagger\rightarrow a^\dagger+h$, together with $B+[a,h]=c$ for constant $c$, gives
\begin{equation}
a(a^\dagger+h)^k\phi_0
=
ck(a^\dagger+h)^{k-1}\phi_0.
\end{equation}
It follows recursively that $a^n\phi_k=0$ for $k=0,\ldots,n-1$. Thus, the scalar components of our multiorbital construction span the same higher-order zero-mode hierarchy familiar from chiral multilayer Hamiltonians~\cite{kailasvuori2009pedestrian}. The distinction is in its realization: here the hierarchy is encoded in internal orbital structure while the Hamiltonian remains first order in momentum, rather than in an $n^{\mathrm{th}}$-order differential operator. 

\subsection{Chiral TBG as a nonuniform magnetic-field problem}

We recall how the exact flat band of chiral twisted bilayer graphene (TBG) may be viewed as a lowest AC band in nonuniform magnetic fields~\cite{ledwith2020fractional}.  The single-valley chiral TBG annihilator takes the form
\begin{equation}
\mathcal{D}_0=
\begin{pmatrix}
        -2\ii \partial_{\bar{z}}&\alpha U(\rr)\\
        \alpha U(-\rr)& -2\ii\partial_{\bar{z}}
    \end{pmatrix},
\end{equation}
where $U(\rr)$ is the periodic moir\'e tunnelling.  At a magic angle, $\mathcal{D}_0$ possesses an exactly flat vortexable zero-mode band~\cite{tarnopolsky2019origin}.  Its wavefunctions may be written as
\begin{equation}
\psi_0(\rr)=f(z)e^{-K(\rr)}\zeta(\rr) = \phi_{0}(\rr)\zeta(\rr),
\end{equation}
where $f(z)$ is holomorphic and $\zeta(\rr)$ is a normalized two-component spinor containing the local layer and phase texture ($\zeta^{\dagger}\zeta=1$).  The scalar envelope $\phi_0(\rr)$ has precisely the standard AC form, and the function $K(\rr)$ defines an effective magnetic field through
$B_{\rm eff}(\rr)=4\partial_z\partial_{\bar z}K(\rr)$.

For the $C=1$ TBG band, the effective field carries one net flux quantum per moir\'e unit cell.  It is therefore useful to separate
\begin{equation}
K(\rr)=\frac{B_0}{4}|z|^2+K_\lambda(\rr),
\end{equation}
where $K_\lambda$ is periodic and
$B_0=\frac{2\pi}{A_{\rm UC}}$.
Thus, the spatial modulation of the TBG wavefunction may be represented by a periodic inhomogeneous magnetic field with positive mean $B_0$.  In the explicit construction of Ref.~\cite{ledwith2020fractional}, this effective field is strongly nonuniform and may even change sign within the moir\'e unit cell.

The correspondence between the AC and TBG formalism can be stated directly at the operator level.  Since $
\mathcal{D}_0\left[e^{-K}\zeta\right]=0$,
the spinor $\zeta$ obeys
\begin{equation}
\mathcal{D}_{0}\zeta
=
-2\ii\left(\partial_{\bar z}K\right)\zeta.
\end{equation}
Consequently, for an arbitrary scalar function $\chi(\rr)$, the first-order nature of $\mathcal{D}_0$ gives
\begin{equation}
\mathcal{D}_0\left[\zeta\chi\right]
=
-2\ii\zeta
\left(\partial_{\bar z}+\partial_{\bar z}K\right)\chi.
\end{equation}
Using the nonuniform-field annihilation operator $a$, this becomes
\begin{equation}
\mathcal{D}_0\left[\zeta\chi\right]
=
\sqrt{2B_0}\,\zeta\,a\chi.
\label{eq:TBG_AC_intertwining}
\end{equation}
Equation~\eqref{eq:TBG_AC_intertwining} is the precise sense in which the chiral TBG annihilator realizes an effective AC problem. The one-dimensional subbundle locally spanned by $\zeta(\rr)$ is preserved by $\mathcal{D}_0$, and the action of $\mathcal{D}_0$ within this subbundle is exactly that of the scalar annihilator $a$.  In particular,
$a\phi_0=0$ implies $\mathcal{D}_0\left[\zeta\phi_0\right]=0$.
The spatially varying effective magnetic field is encoded in the scalar envelope $e^{-K}$, while the remaining microscopic information about the TBG wavefunction is carried by the spinor $\zeta(\rr)$. Hence, any operator built from copies of $\mathcal{D}_0$ and scalar inter-copy couplings induces, on the corresponding vortexable subbundles, an operator built from copies of the nonuniform-field annihilator $a$.

\end{document}